\documentclass[10pt,a4paper]{article}
\usepackage[utf8]{inputenc}
\usepackage{amsmath}
\usepackage{amsfonts}
\usepackage{amssymb}
\usepackage{eurosym}
\usepackage{graphicx}
\usepackage{hyperref}
\usepackage{comment}
\usepackage[backend=biber,bibencoding=utf8, style=nature]{biblatex}
\usepackage[font={small,it}]{caption}
\usepackage{booktabs} 
\usepackage{makecell}
\usepackage{amsmath}
\usepackage{multirow}
\DeclareMathOperator*{\argmax}{arg\,max}
\usepackage{gensymb}
\usepackage{geometry}
\usepackage[backend=biber,bibencoding=utf8, style=nature]{biblatex}
\usepackage{authblk}
\title{Superintense Laser-driven Photon Activation Analysis}
\author[1]{F. Mirani\footnote{e-mail: francesco.mirani@polimi.it }}
\author[2]{D. Calzolari}
\author[1]{A. Formenti}
\author[1]{M. Passoni}
\affil[1]{\emph{Politecnico di Milano, Via Ponzio 34/3, I-20133 Milan, Italy}}
\affil[2]{\emph{CERN, Geneva, Switzerland}}
\date{February 2021}                     
\begin{document}
\maketitle

\begin{abstract}
Laser-driven radiation sources are attracting increasing attention for several materials science applications. While laser-driven ions, electrons and neutrons have already been considered to carry out the elemental characterization of materials, the possibility to exploit high energy photons remains unexplored. Indeed, the electrons generated by the interaction of an ultra-intense laser pulse with a near-critical material can be turned into high energy photons via bremsstrahlung emission when shot into a high-Z converter. These photons could be effectively exploited to perform Photon Activation Analysis (PAA). In the present work, the possibility to perform laser-driven PAA is proposed and investigated. By means of a theoretical and numerical approach, we identify the optimal near-critical material and converter parameters for laser-driven PAA in a wide range of laser intensities. Finally, exploiting the Monte Carlo and Particle-In-Cell tools, we simulate PAA experiments performed with both conventional accelerators and laser-driven sources.
\end{abstract}

\section{Introduction} \label{into_sect}
\begin{figure}[b!]
\centerline {\includegraphics[scale=0.38]{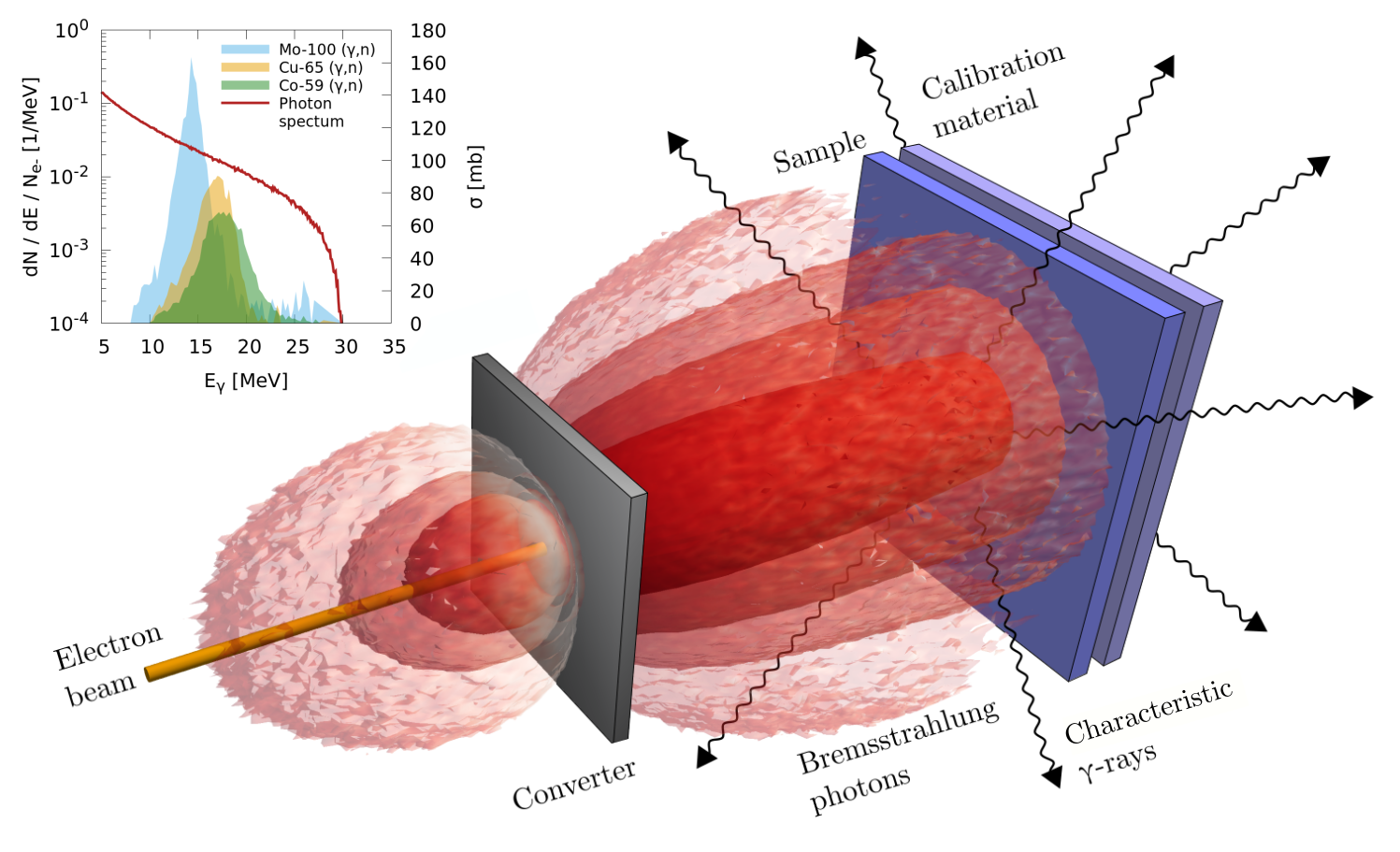}}
\caption{Conceptual setup of Photon Activation Analysis. The inset graph reports some examples of $(\gamma, n)$ photonuclear cross sections (from the Evaluated Nuclear Data File (ENDF) database \cite{brown2018endf}). The red line is the bremsstrahlung photon energy spectrum obtained from a Fluka simulation performed with 30 MeV monoenergetic electrons and a W converter having thickness equal to 3 mm.}
\label{Intro_imm}
\end{figure}

Photon Activation Analysis (PAA) \cite{segebade2011PAA_book, segebade2006PAA_rew} is a non-destructive materials characterization technique which exploits high-energy photons to retrieve the elemental composition of a large variety of samples.  During the irradiation time, the photons interact with both the sample under study, inducing photonuclear reactions within these targets. After a cooling period (i.e. rest time), delayed characteristic $\gamma$-rays are emitted and detected. These $\gamma$-rays are exploited to reconstruct the sample composition. From the knowledge of the characteristic $\gamma$-ray energy, the nuclear reaction and the decay channel, the parent element is identified. Moreover, taking into account the number of $\gamma$-rays, the concentrations of the elements are reconstructed as well. Often, multiple measurements are performed after stepwise increasing the rest time to avoid spectral interference by short-lived nuclides. Because several physical quantities (e.g. the nuclear cross sections) and experimental parameters are not well known a priori, the elemental concentration reconstruction is performed exploiting a comparative (or calibration) material of known composition, which is co-irradiated with the sample. The primary photons are generated through the interaction of a high energy (e.g. $20-30$ MeV) electron beam with a mm thick converter material having high atomic number (e.g. tungsten). The primary electron beam is provided by a high power linear accelerator, microtron or betatron with delivered current of $\sim 10$ $\mathrm{\mu}$A. Exploiting bremsstrahlung (BS) emission, about 50$\%$ of the primary electron energy is converted into photons whose energies range from zero to that of the primary electrons \cite{starovoitova2016sources}. The conventional PAA irradiation setup, some examples of photonuclear cross sections and a BS energy spectrum are reported in figure \ref{Intro_imm}. Nowadays, PAA is routinely exploited for environmental \cite{masumoto1999envirorment}, biological \cite{kato1976biological}, geochemical \cite{vranda2007biology}, archaeological \cite{reimers1977archeology} and industrial studies \cite{leonhardt1982coal}. \\
Despite the great PAA analytical capabilities, the employed electron accelerators are characterized by large dimensions and costs, with strong limitations to the widespread use of this technique. Laser-driven particle sources \cite{macchi2013Laser_review, daido2012Laser_review, corde2013femtosecond, esarey2009physics, alejo2015recent} may represent a promising alternative to conventional accelerators. They rely on the interaction of a super-intense ($I > 10^{18} \mathrm{W/cm^{2}}$) and ultra-fast laser pulse (few J, $\sim10$ fs) with a target to generate high energy particles. The collectively accelerated particles are characterized by an ultra-fast dynamics and broad energy spectra. Interestingly, the generated particles type and properties can be tuned by acting on the laser parameters and target configuration. As a consequence, these sources are potential multifunctional tools for several applications in materials and nuclear science \cite{passoni2019Mat_Nuc}. For instance, laser-driven protons and electrons accelerated exploiting 10-100 TW class lasers and micrometric solid foils can be exploited for Particle Induced X-ray Emission (PIXE) \cite{barberio2017laser, passoni2019superintense, mirani2020PIXE_EDX} and Energy Dispersive X-ray (EDX) spectroscopy \cite{mirani2020PIXE_EDX}. In addition, the adoption of advanced Double Layer Targets (DLT) \cite{passoni2016foam_1, prencipe2016development, passoni2014energetic}, where the thin foil is coated with a near-critical carbon foam \cite{zani2013foam}, is a valid strategy to increase the number and energy of the accelerated electrons and ions. \\
In recent works \cite{rosmej2019interaction, rosmej2020high, willingale2018unexpected}, exploiting a near-critical targets and $\sim 100$ TW class lasers, even if using quite different laser drivers with longer pulses and higher energies (~100 fs, ~100 J), the possibility to accelerate electron bunches with maximum energy up to 80 MeV and total accelerated charge of several $\mathrm{\mu}$C was demonstrated. An alternative, extensively investigated strategy for high energy electrons production is to use under-critical gas-jet targets and the wakefield acceleration (lwfa) mechanism \cite{malka2008wakefield}. In this case, exploiting $\sim10-100$ TW lasers working at $1-10$ Hz repetition rate, the accelerated electrons are quasi-monoenergetic with energies up to $\sim \mathrm{GeV}$ and accelerated charge of $\sim 10-100$ pC per bunch \cite{leemans2006gev, kim2013enhancement, gonsalves2019petawatt, maier2020decoding}. In addition, sub-TW class lasers (i.e. 10s mJ of energy) have been successfully exploded to perform lwfa at kHz repetition rate \cite{gustas2018high}. The energy of the electrons is of the order of $\sim 10$ MeV with delivered currents of $\sim 10$ nA. The conversion of lwfa electrons into bremsstrahlung radiation for applications \cite{ledingham2010laser} like imaging \cite{ben2011compact}, neutron production \cite{galy2007bremsstrahlung} and integral ($\gamma$ ,n) cross-sections measurements \cite{ledingham1999nuclear} has been proposed. \\
On the other hand, the possibility of performing PAA with photons generated from laser-driven electrons has not been considered yet. The main goal of this work is to numerically investigate this potential application of laser-driven electron sources. To this aim, even if both lwfa and near-critical solid targets are worth of consideration, we focus on the second approach because of the higher delivered charge per laser shot. We consider a low density material, e.g. a foam, for the electron acceleration, attached to a mm thick W plate for the production of high energy photons. Note that this peculiar DLT configuration is compatible with the experimental setup for ion acceleration based on thin solid targets. Therefore, another advantage over lwfa consists in the possibility of switching from proton acceleration to photon production with minimum changes to the experimental apparatus.
\\This work has two objectives. The first one consists in the development of a theoretical description of PAA, performed with either conventional or laser-driven sources. This was achieved exploiting Fluka \cite{ferrari2005fluka} Monte Carlo (MC) simulations and a proper theoretical description of laser-driven electrons acceleration in near-critical density media \cite{pazzaglia2020theoretical}. This theoretical description allows us to optimize the laser-driven source parameters to achieve the maximum PAA performances. Moreover, the model is used to compare the laser-driven PAA capabilities with that accomplished exploiting conventional accelerators. The second goal of the work is to simulate specific PAA experiments by means of the MC tool. The simulations are carried out considering both monoenergetic and broad laser-driven like electron energy spectra. Moreover, the latter are described both with a simplified analytical description and performing a 3D Particle-In-Cell (PIC) \cite{birdsall2004plasma, arber2015contemporary}. We provide the electron momenta distribution from the PIC as input to the MC simulation. This analysis shows that the model introduced in the first part of the work provides reliable results and that laser-driven sources may be effectively exploited for PAA studies.

\section{Results and discussion}
\subsection{Theoretical models for conventional and laser-driven PAA}
We start considering monoenergetic electrons and we perform a parametric scan to determine the optimal converter thickness as a function of the incident electron energy (i.e. the thickness that maximizes the number of emitted characteristics $\gamma$-ray from the sample). Then, we carry out the same scan considering electrons with exponential energy spectra, thus compatible with the electrons provided by a laser-driven source. Exploiting the model available from \textit{Pazzaglia et al.} \cite{pazzaglia2020theoretical}, we introduce the main laser and target parameters in the description. The goal is to find the optimal target parameters (i.e. near-critical layer density and thickness) for the laser-driven electron source. Moreover, we want to compare the laser-driven PAA performances, in terms of the emitted characteristic $\gamma$-ray signals, with that of the conventional accelerators.
\subsubsection{Model for PAA with monoenergetic electrons}\label{sec_model_PAA_mono}
\begin{figure}[t!]
\centerline {\includegraphics[scale=0.50]{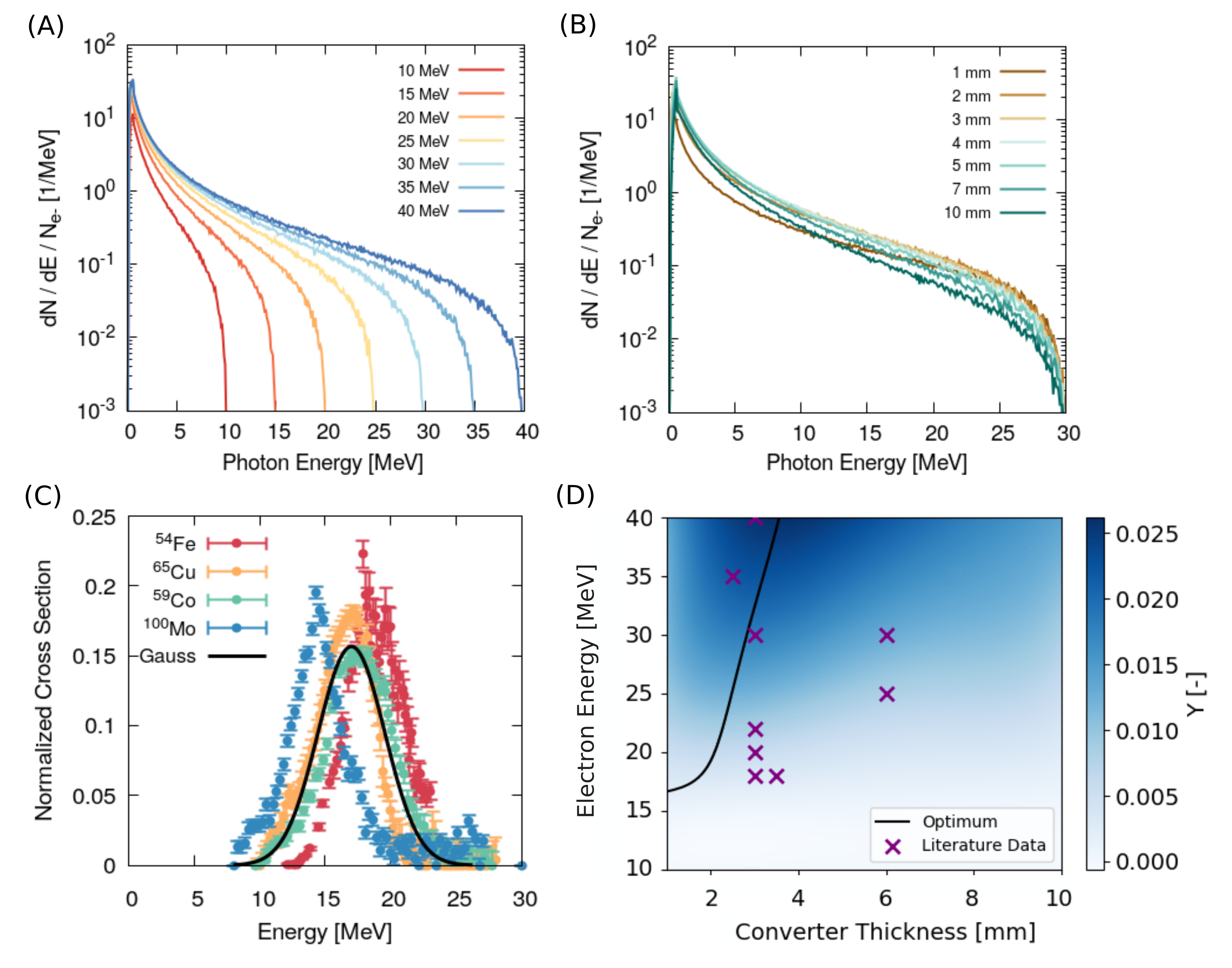}}
\caption{(A) Photon energy spectra from Fluka simulations for different values of the primary electron energy and fixed converter thickness (i.e. 3 mm). (B) Photon energy spectra from Fluka simulations for different values of the converter thickness and fixed primary electron energy (i.e. 30 MeV). The spectra are normalized to the total number of primary electrons. (C) Comparison between a Gaussian function centred in 17 MeV, $FWHM = 7$ MeV and normalized experimental photonuclear cross sections. (D) Heat-map showing the normalized yield as a function of the primary electron energy and converter thickness. The black line marks the locus of optimal converter thickness, while the markers correspond to working points from literature \cite{krausova2015nondestructive, eke2016determination, vranda2003elemental, randa2007instrumental, starovoitova2016high, aliev2005use, hislop1972determination, landsberger1985analysis}.}
\label{Mono_scan}
\end{figure}

Two fundamental ingredients for the evaluation of the PAA performances are the BS production of $\gamma$-rays and the probability to have photonuclear reactions.
As far as the BS emission is concerned, the generated photon spectrum $f(E) = (1/N_{e}) dN_{\gamma}/dE$ per unit of incident electron $N_{e}$ can be obtained by means of analytical formulas \cite{findlay1989analytic} or MC simulations \cite{berger1970bremsstrahlung}. The first approach is very useful to provide quick estimations. However, it requires the adoption of simplifying assumptions which, in some cases, can lead to a non-acceptable degree of error. Accordingly, a MC description of the process is often demanded for several applications, as in the case of PAA studies \cite{starovoitova2016high}. Therefore, we performed a set of 49 MC simulations to evaluate the BS energy spectra obtained from different primary electron energies (i.e. 10, 15, 20, 25, 30, 35 and 40 MeV) and tungsten converter thicknesses (i.e. 1, 2, 3, 4, 5, 7, 10 mm). The selected values cover the ranges usually exploited in PAA experiments. Some examples of the obtained $\gamma$-ray spectra are reported in figure \ref{Mono_scan}(A) and \ref{Mono_scan}(B) for different primary electron energies and converter thicknesses, respectively.\\
Even if $(\gamma, 2n)$, $(\gamma, 3n)$, $(\gamma, p)$ reactions can take place, the $(\gamma, n)$ are the most commonly exploited for PAA purposes \cite{segebade2006PAA_rew}. Therefore, that is the only kind of photonuclear events considered here. The cross sections are well-described by a bell-shape function centred approximately around 15-20 MeV (see figure \ref{Intro_imm}(B)). The position of the maximum is weakly dependent on the isotope and its magnitude can range from few tens up to hundreds of mbarn. We model the cross section as $\sigma(E) = \sigma_{int} \widetilde{\sigma}(E)$, where $\sigma_{int}$ is the total area and $\widetilde{\sigma}(E)$ is a normalized cross section described as a Gaussian function. A comparison between the $\widetilde{\sigma}(E)$ and experimental cross sections upon normalization is shown in figure \ref{Mono_scan}(C). To generalize the discussion, we will exploit $\widetilde{\sigma}(E)$, instead of $\sigma(E)$, in the remaining part of the work.\\
We evaluate the integral:
\begin{equation} \label{eq:activation_prop}
Y(E_{e},l) = \int_{E}f(E)\times\widetilde{\sigma}(E)dE
\end{equation} 
considering the BS spectra $f(E)$ obtained from the MC simulations. $Y(E_{e},l)$ is the normalized reaction yields (i.e. per unit of electron) for the different primary electron energies $E_{e}$ and converter thicknesses $l$. To retrieve the value of $Y(E_{e},l)$ for the whole range of electron energies and thicknesses, we fit the discrete values with a fourth order polynomial in the variables $E_{e}$ and $l$. The resulting continuous function $Y(E_{e},l)$ is represented as blue color map in figure \ref{Mono_scan}(D). Finally, we evaluate the optimal converter thickness $t_{opt}$ as a function of the incident electron energy from:
\begin{equation} \label{eq:vect_potent}
l_{opt} (E_{e}) = \argmax_{l} Y (E_{e},l)
\end{equation}
and the corresponding values of normalized yields $Y_{max} (E_{e}) = Y(E_{e}, l_{opt}(E_{e}))$. The function $l_{opt} (E_{e})$ is superimposed to the heat map in figure \ref{Mono_scan}(D). In addition, we marked the points corresponding to some literature experimental conditions adopted in the past for PAA. We want to point out that the electron current $I_{e}$, which is an important operating parameter for PAA, is not considered here. This is due to the fact that the electron energy and current are considered as independent parameters and the optimal converter thickness can be obtained without taking into account the latter.\\
From the obtained results, we observe that the performances of PAA, expressed in terms of reaction yields, are weakly dependent on the adopted converter thickness. The optimum is between 2 and 4 mm and several converter thicknesses from literature are in this range. On the other hand, the dependence on the electron energy is stronger. Below 15 MeV electron energy, the yield tends to zero because the continuum BS spectrum does not cover the region where the cross section is located. Accordingly, the curve for the optimal thickness asymptotically tends to zero around $\sim 15$ MeV of energy. Below this value, no BS photons useful for PAA are produced, whatever the value of the converter thickness.

\subsubsection{Model for PAA with laser-driven electrons}\label{sec_model_PAA_laser}
\begin{figure}[b!]
\centerline {\includegraphics[scale=0.45]{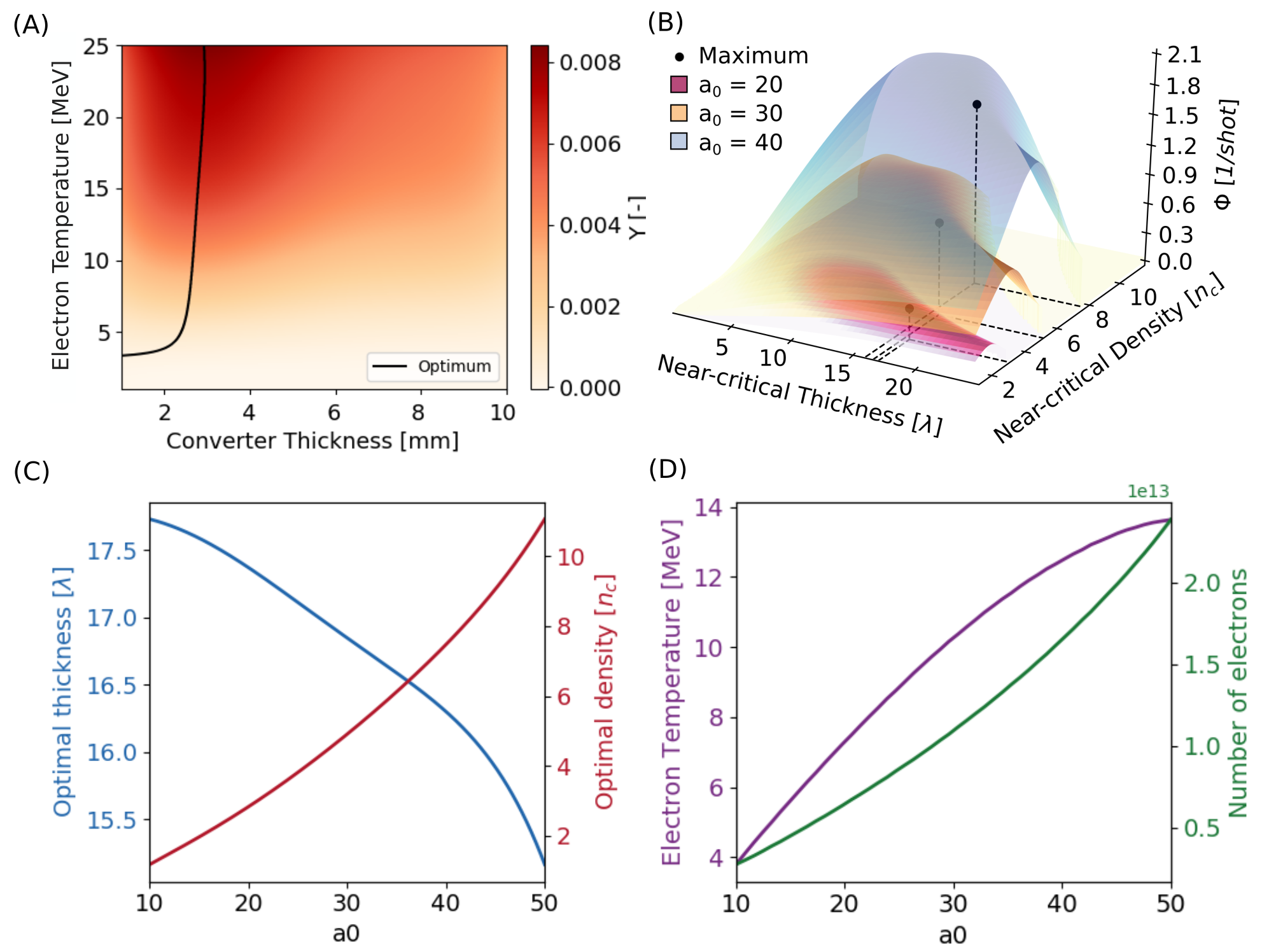}}
\caption{(A) Heat-map showing the normalized yield as a function of the laser-driven electron temperature and converter thickness. The black line marks the locus of optimal converter thickness. (B) Normalized yield per laser shot for three values of $a_{0} = 20$ (i.e. 20, 30 and 40) as a function of the target near-critical layer density and thickness. The positions of the maxima are marked with black dots. (C) Trends for the optimal near-critical layer thickness and density as a function of $a_{0}$. (D) Trends for the electron temperature and number of accelerated electrons per shot as a function of $a_{0}$.}
\label{Exp_scan}
\end{figure}

We consider laser-driven electrons characterized by exponential energy spectra with maximum energy equal to 40 MeV. With this choice of the cut-off, the electron energies, and therefore the generated BS spectra, fall in the region of non-vanishing photonuclear cross sections. The value is quite reasonable for sufficiently intense laser pulses (normalized laser intensity $a_{0} \ge 10$) and targets coated with a near-critical density layer \cite{bin2018enhanced}. 
However, since the electron spectrum is exponential, very few BS photons are produced in correspondence with the cut-off. Therefore, the exact value of the maximum electron energy has a negligible effect on the number of activated nuclei.\\
We carry out 35 MC simulations to retrieve the BS energy spectra for various electron temperatures $T_{e}$ (i.e. 5, 10, 15, 20 and 25 MeV) and converter thicknesses (i.e. 1, 2, 3, 4, 5, 7 and 10 mm). We exploit the same procedure described in the previous section to find the normalized reaction yields for each value of $T_{e}$ and $l$, as well as the optimal converter thickness $l_{opt}(T_{e})$ (see figure \ref{Exp_scan}(A)).
\\In order to proceed, the maximum normalized yield $Y_{max}(T_{e})$ (i.e. the values corresponding to the optimal laser-driven PAA conditions) must be expressed in terms of the laser and target operating parameters. Here we consider the near-critical layer thickness $r_{nc}$ and density $n_{nc}$ of the target and the normalized laser intensity $a_{0}$. To this aim, we exploit the recently developed theoretical model from \textit{Pazzaglia et al.} \cite{pazzaglia2020theoretical} (see the Methods section for details). The model allows us to express the aforementioned operating parameters in terms of the electron temperature and, therefore, to $Y_{max}$. In this case, the number of accelerated electrons per shot $N_{e}$ is not decoupled from the temperature. Indeed, both $T_e$ and $N_e$ depend on the model parameters, here $r_{nc}$, $n_{nc}$ and $a_{0}$. Therefore, the definition of the optimal laser-driven PAA conditions needs to be extended, including the number of accelerated electrons per shot. The quantity that will be further maximized is the normalized yield per laser shot:
\begin{equation} \label{eq:rate_act_laser}
\Phi(r_{nc},n_{nc},a_{0})=Y_{max}(T_{e}(r_{nc},n_{nc},a_{0}))\times N_{e}(r_{nc},n_{nc},a_{0})
\end{equation}
where $T_{e}$ and $N_{e}$ are related to $r_{nc}$, $n_{nc}$ and $a_{0}$ according to the aforementioned model. In the model, we assume a linearly polarized laser pulse with normal incidence, wavelength $\lambda = 0.8$ $\mathrm{\mu}$m and waist $FWHM = 4.7$ $\mathrm{\mu}$m. The curves in figure \ref{Exp_scan}(B) show the behaviour of $\Phi$ as a function of $r_{nc}$ and $n_{nc}$ for three values of $a_{0}$. Then, for a certain value of $a_{0}$, the optimal near-critical thickness and density are obtained from:
\begin{equation} \label{eq:rate_act_laser_max}
r_{opt}(a_{0}), n_{opt}(a_{0}) = \argmax_{n_{nc}, r_{nc}}\Phi(r_{nc},n_{nc},a_{0})
\end{equation}
The corresponding values are marked in figure \ref{Exp_scan}(B) as well.\\
We evaluate equation \ref{eq:rate_act_laser_max} for $a_{0}$ ranging from 10 to 50. The resulting optimal target parameters are reported in figure \ref{Exp_scan}(C). The thickness and density are in units of $\lambda$ and critical density $n_{c}$, respectively. For each value of $a_{0}$, the reported $r_{opt}$ and $n_{opt}$ maximize the activation of the sample. That maximization is the result of a trade off between the accelerated electron temperature and number, which in turn affect the BS photon spectrum. The values of $T_{e,opt}(a_{0})$ and $N_{e,opt}(a_{0})$ are reported in figure \ref{Exp_scan}(D). As expected, they monotonically increase with the intensity of the laser. We can also obtain the corresponding values for the normalized yield $\Phi_{max}(r_{opt}(a_{0}), n_{opt}(a_{0}),a_{0})$.

\subsubsection{Comparison between conventional and laser-driven electron sources} \label{comp_section}
In sections \ref{sec_model_PAA_mono} and \ref{sec_model_PAA_laser}, we have identified the optimal parameters for conventional PAA (i.e. the converter thickness as a function of the electrons energy) and for laser-driven PAA (i.e. the converter thickness, the near-critical layer thickness and density as a function of the laser intensity). Now, the aim is to compare the performances of the PAA, in terms of sample activation rates, carried out with monoenergetic and laser-driven electron sources. For this reason, the normalized yields per impinging electrons $Y_{max}(E_{e})$ must be multiplied by the electron current $I_{e}$ provided by the conventional accelerator. As far as the laser-driven source is concerned, we consider the product between the normalized yield per laser shot $\Phi_{max}(a_{0})$ and the repetition rate $RR$.
\begin{figure}[b!]
\centerline {\includegraphics[scale=0.50]{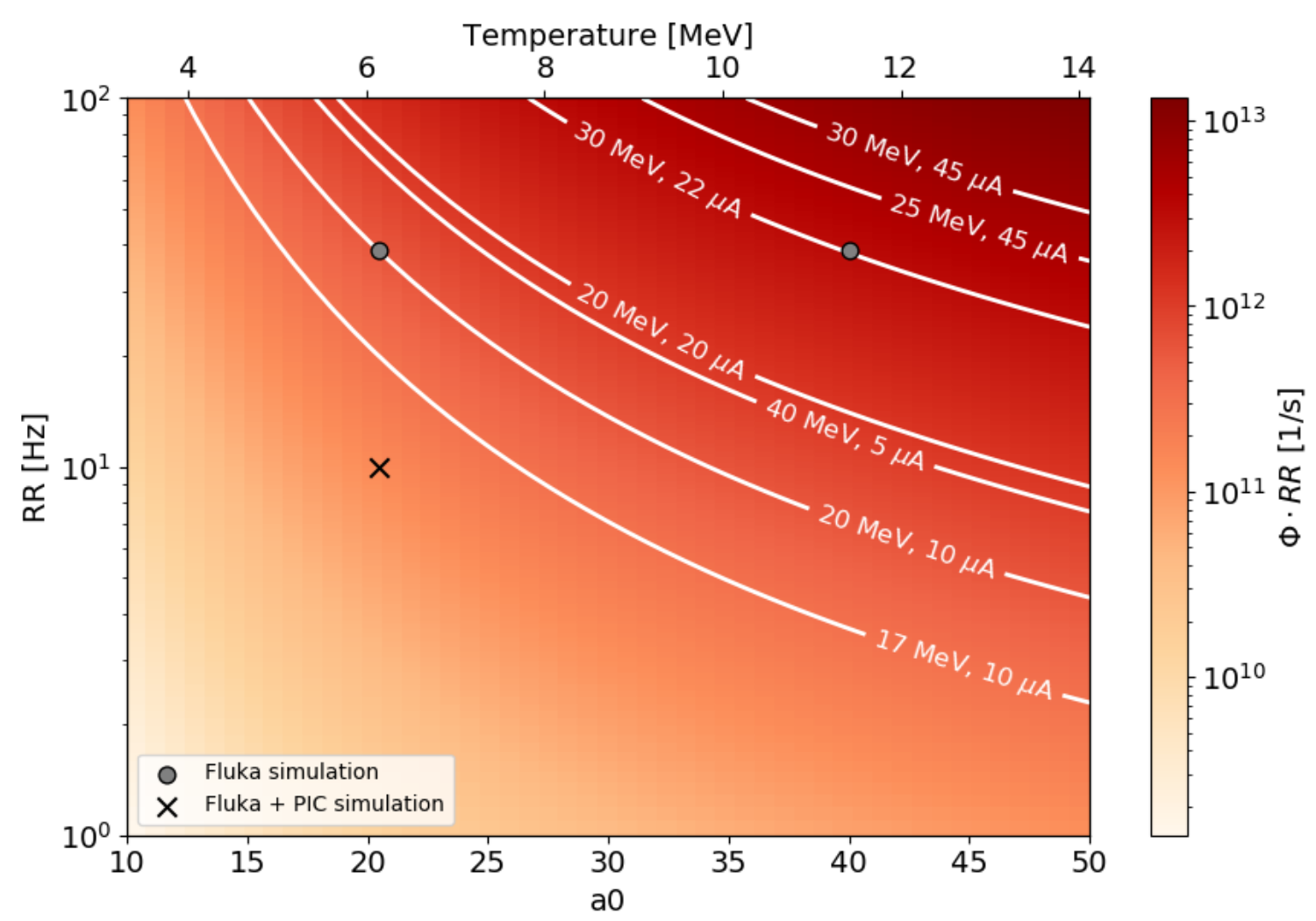}}
\caption{Comparison between laser-driven and monoenergetic sources for PAA. The red heat-map represents the normalized activation rate obtained for laser-driven sources. The isolines correspond to the normalized activation rate that can be obtained with some existing conventional accelerators for PAA (reported electron energies and currents from literature \cite{krausova2015nondestructive, randa2007instrumental}). The grey points identify the two selected experimental conditions for which the full PAA Monte Carlo simulations are performed. The cross marker identifies the laser parameters for the realistic laser-driven PAA simulation.}
\label{comparison_final}
\end{figure}
\\In figure \ref{comparison_final}, the behaviour of the quantity $\Phi_{max} RR$ is shown as red heat-map for $a_{0}$ and $RR$ in the ranges $10 - 50$ and $1 - 100$ Hz, respectively. On the upper x-axis, we report the corresponding values of $T_{e,opt}(a_{0})$ obtained in the previous section. The isolines represent specific monoenergetic sources identified by the energy and current of the primary electrons. Their positions are fixed by the relation $\Phi_{max}(a_{0}) RR = Y_{max}(E_{e}) I_{e}$. Therefore, each isoline for the mononenergetic source lies in correspondence with several possible equivalent laser systems. The reliability of the comparison will be carefully checked in the following section.  Here, we aim at discussing the feasibility of laser-driven PAA.\\
In order to generate enough BS photons to sufficiently activate the sample, both the laser intensity (thus the electron temperature) and the repetition rate must be taken into account. Indeed, the performances of laser-driven PAA will be given by a trade-off between these two parameters. The existing ultra-intense lasers already provide intensities in the entire range considered for $a_{0}$. However, the nominal repetition rate that they can currently reach is of $1 - 10$ Hz \cite{danson2019petawatt}. Therefore, the upper region of the map in figure \ref{comparison_final} (i.e. where some the most powerful conventional accelerators for PAA are located) is not achievable yet. Nevertheless, considering the lower region of the map, we observe that 100s TW class laser and near-critical targets can approach the performances of, at least, some conventional accelerators for PAA.\\
We also want to point out that the research in ultra-intense laser technology is very active. In this respect, a further improvement in the performances and a reduction of dimensions and costs are foreseen in the following years. Thus, laser-driven electron sources could be on their way to become competitive with conventional accelerators for applications like PAA. 

\subsection{PAA simulations with monoenergetic and laser-driven electrons}
The second part of this work is focused on the Fluka MC simulation of PAA experiments performed with both monoenergetic and laser-driven sources. The simulations take into account the generation of BS photons, the activation of the sample and standard material, the decay and delayed emission of characteristic $\gamma$-rays. A detailed description of the MC simulations is provided in the Methods section. \\
We consider 5 experimental conditions, two of which (called $S1$ and $S2$ from here on) involve monoenergetic electron sources with energies of 30 and 22 MeV and currents equal to 45 and 15 $\mu$A, respectively. For these two cases, we simulate the equivalent PAA experiments (i.e. $S3$ and $S4$) performed with laser-driven sources. Accordingly to the model described in the previous Section, they should provide the same values of normalized activation rate. The considered experimental conditions are marked as grey points in figure \ref{comparison_final}. The laser-driven electron energy spectra are modeled as pure exponential functions with cut-off energies of 40 MeV. The goal is to compare the number of emitted characteristic $\gamma$-rays for all four cases and check the reliability of the comparison in section \ref{comp_section}. The last MC simulation (i.e. $S5$) is related to a laser-driven PAA experiment performed with realistic laser parameters (the cross marker in figure \ref{comparison_final}). In this case, the electron energy spectrum and angular divergence are obtained from a 3D Particle-In-Cell simulation. The aim is to asses whether the elemental composition of the sample can be obtained with less demanding laser requirements.\\
The sample and standard compositions and dimensions are the same for all the simulations. They are compatible with the content of a bronze sculpture (1550–1400 BC) analyzed by Prompt Gamma Activation analysis and Neutron
Imaging \cite{maroti2017characterization}. The sample and standard contain several elements (see the Methods section). As far as the activation is concerned, we focus on Cu, Na, Fe, Pb, Ni and Ca.  For each PAA simulated experiment, we consider three different cooling times and measurement times. A summary of the main parameters for all the simulations is provided in table \ref{Summ_parameters}. 
\begingroup
\setlength{\tabcolsep}{6pt} 
\renewcommand{\arraystretch}{1.2} 
\begin{table}[h!]
\caption{Summary of the parameters adopted in the simulations.}
\centering
\footnotesize
\begin{tabular}{cccccc}
\toprule
Source&\multicolumn{2}{c}{\makecell[c]{Monoenergetic}}&\multicolumn{3}{c}{Laser-driven}\\
\midrule
Simulation&$S1$&$S2$&$S3$&$S4$&$S5$\\
\midrule
Electrons energy ($E_{e}$) [MeV]&30&22&-&-&-\\
Electron current ($I$) [$\mu A$]&45&15&-&-&-\\
Laser intensity ($a_0$) [-]&-&-&40&20.5&20.5\\
Laser repetition rate ($RR$) [Hz]&-&-&38.6&38.6&10\\
Near-critical density ($n_{opt}$) [$n_c$]&-&-&7.5&2.92&2.92\\
Near-critical thickness ($r_{opt}$) [$\lambda$]&-&-&16.3&17.34&17.34\\
Electrons temperature ($T_{e}$) [MeV]&-&-&12.5&7.5&PIC\\
Electrons per shot ($N_{e}$) [-]&-&-&$1.6\times 10^{13}$&$6.4\times 10^{12}$&PIC\\
Converter thickness ($t$) [mm]&2.41&3&2.73&2.58&2.58\\
Irradiation time ($t_{i}$) [h]&3&3&3&3&3\\
Rest times ($t_{r}$) [d]&0.5, 7, 30&0.5, 7, 30&0.5, 7, 30&0.5, 7, 30&0.5, 7, 30\\
Measurement times ($t_{m}$) [h]&2, 8, 24&2, 8, 24&2, 8, 24&2, 8, 24&2, 8, 24\\
\bottomrule
\end{tabular}\label{Summ_parameters}
\end{table}
\endgroup

\begin{figure}[h!]
\centerline {\includegraphics[scale=0.50]{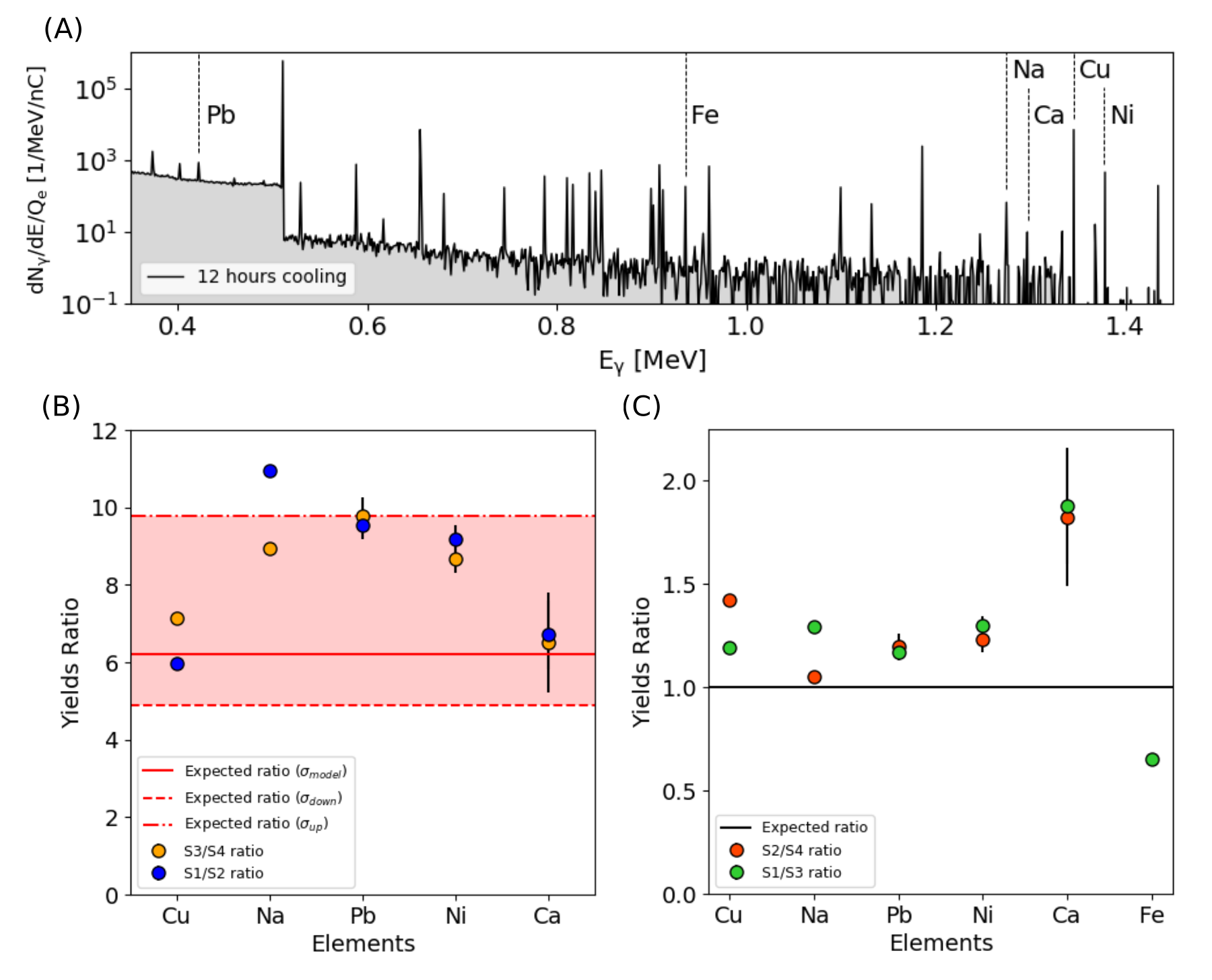}}
\caption{(A) Photon energy spectrum from Monte Carlo simulation $S1$ obtained from sample irradiation after 12 hours rest time and for 3 hours measurement time. The considered characteristic $\gamma$-ray peaks for the elements are identified by the dashed lines. (B) Ratio between the intensities of simulations $S1/S2$ and $S3/S4$ for each element. The continuous red line corresponds to the ratio predicted by the model assuming a normalized cross section centered at 17 MeV. The dashed upper and lower red lines correspond to the ratio predicted by the model assuming normalized cross sections centered in 20 MeV and 15 MeV, respectively. (C) Ratio between the intensities of simulations $S1/S3$ and $S2/S4$ for each element. The black line corresponds to the value predicted by the model. 
}
\label{comparison_final_sim}
\end{figure}

\subsubsection{Test of the comparison between conventional and laser-driven PAA}
From the MC simulations, we retrieve the characteristic $\gamma$-rays energy spectra collected during the measurement times, following the irradiation and rest times. An example of spectrum, from the $S1$ simulation and recorded after a rest time of 12 hours, is shown in figure \ref{comparison_final_sim}(A). The positions of the considered characteristic peaks for the various elements are highlighted. To assess the feasibility of the comparison presented in Section \ref{comp_section}, we perform the ratio between the peak intensities obtained in the simulations $S1$, $S2$, $S3$ and $S4$ for the various elements. Then, we compare the simulated ratios with that predicted by the theoretical model. We perform the intensity ratios because the results are independent from several parameters like the sample mass, thickness and elemental concentrations, the specific values of $\widetilde{\sigma}$, the irradiation, measurement and cooling times. Therefore, the ratios between line intensities (from the MC) can be directly compared with the ratios between normalized activation rates (from the theoretical model).\\
We start considering the ratios between the intensities of simulations $S1/S2$ (i.e. the monoenergetic electron sources) and $S3/S4$ (i.e. the laser-driven electron sources). They are plotted in figure \ref{comparison_final_sim}(B) as blue and yellow points for all the elements. The ratios involve simulations performed in correspondence of different points of the map in figure \ref{comparison_final}. The predicted result is 6.2 considering a normalized cross section $\widetilde{\sigma}(E)$ centered in 17 MeV (the continuous red line in figure \ref{comparison_final_sim}(B)). Assuming a cross section centered in 15 MeV and 20 MeV, the expected ratio results 4.9 and 9.8, respectively (the dashed lines in figure \ref{comparison_final_sim}(B)). Clearly, the expected ratio, and therefore the predicted performances of laser-driven PAA, depends on the choice of the parameters for the cross section. This is also confirmed by the different ratios obtained for the various elements. Nevertheless, all the points lie within the interval identified by the lower and upper values adopted for the definition of the cross section. Therefore, for both monoenergetic and laser-driven electrons, the model satisfactorily predicts how the characteristic peak intensities scale for sources with different operating parameters.\\
In figure \ref{comparison_final_sim}(C), the ratio between the intensities of simulations $S1/S3$ and $S2/S4$ (i.e. the monoenergetic sources over the equivalent laser-driven ones) are shown. In this case, the expected ratio is equal to 1 (the continuous black line). With the only exception of Ca, all the points lie in a region close to the expected value. On average, the discrepancy is of the order of 20\%. Therefore, the comparison allows us to establish an equivalence between monoenergetic and laser-driven sources within an acceptable range of reliability.\\
Finally, it is worth mentioning that the points associated to the Fe peak are not present in figure \ref{comparison_final_sim}(B). This is due to the fact that the generated BS photons in simulations $S2$ and $S4$ are not energetic enough to induce sufficient ($\gamma$, n) reactions.  

\subsubsection{Combined PIC and Monte Carlo simulation of laser-driven PAA}
\begin{figure}[b!]
\centerline {\includegraphics[scale=0.25]{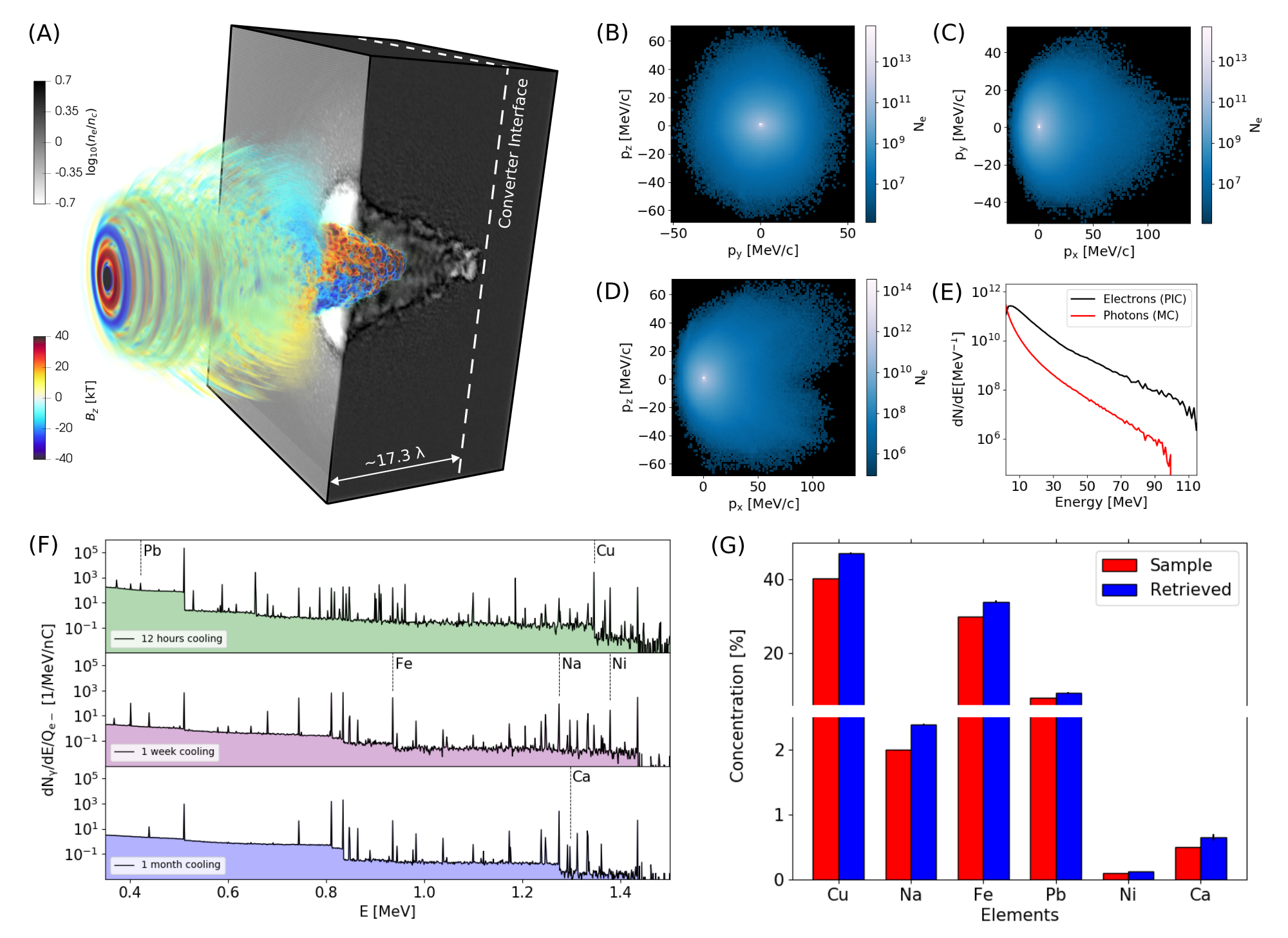}}
\caption{(A) Exploded view of the electron density and z-component of the magnetic field from the PIC simulation. They are retrieved at the instant of time in which the laser front is in correspondence of the rear side (the dashed white line) of the near-critical layer. (B-D) Momenta distributions of the electrons from the PIC simulation. The distributions are integrated along one direction in phase space and plotted in the other two. (E) Electron energy spectrum from PIC and BS photon spectrum from MC. (F) Simulated photon energy spectra emitted following the activation of the sample. They are recorded after three different cooling times. (G) Comparison between the retrieved elemental concentrations and the values set in the MC simulation.}
\label{laser_driven_PAA_sim}
\end{figure}
The last laser-driven PAA simulation (i.e. S5) involves laser parameters compatible with the state of the art technology, while the near-critical target and converter properties are optimized according to the results presented in Sections \ref{sec_model_PAA_laser}. The main parameters are summarized in the last column of Table \ref{Summ_parameters}. As already mentioned in the \ref{into_sect}, we consider a two-layer structure where the near-critical medium is attached to the high-Z material. For our purposes, this choice has several advantages. \\
Electrons accelerated by the laser in the first layer are directly injected inside the converter. The main advantage is to avoid geometrical losses due to the presence of a vacuum gap between the target and converter. Secondly, the establishing of the TNSA process is not allowed. Indeed, in TNSA, a small part of the electron energy would be transferred to ions with a modest reduction of photon generation due to electron BS. Moreover, the proposed configuration is very robust from the structural point of view. Compared to DLTs made by thin substrates, a mm thick target does not require the encapsulation in a perforated holder. The laser can be moved and focused quickly at any point on the surface aiding the high repetition rate operation. Finally, such target-converter artifact can be easily manufactured. For instance, a near-critical carbon foam could be directly deposited on the surface of a tungsten plate via Pulsed Laser Deposition \cite{zani2013foam}. \\
In order to simulate the electron generation in the near-critical layer, we perform a 3D PIC simulation. Technical details are provided in the Methods Section. The presence of the dense converter would require unbearable computational resources and it can not be included in the simulation. Therefore, we consider a thicker medium with respect to the actual near-critical layer to avoid electron expansion in vacuum at the rear side of the medium. Figure \ref{laser_driven_PAA_sim}(A) shows the electron density and the z-component of the magnetic field at the instant in which the laser front reaches the near-critical layer thickness we intend to consider (i.e. $17.34 \mathrm{\lambda}$). At this instant of time (i.e. 166 fs after the start of the simulation), we retrieve the electron momenta distribution $r^{3}N/dp_{x}dp_{y}dp{z}$ from the macro-electron phase space and we do not consider further propagation. The $d^{2}N/dp_{i}dp_{j}$ distributions integrated along the missing component are shown in figures \ref{laser_driven_PAA_sim}(B-D). Clearly, electrons are characterized by a symmetric distribution in the $p_{y} - p_{z}$ plane, while they are forward peaked along the $p_{x}$ component (which corresponds to the laser propagation direction). Figure \ref{laser_driven_PAA_sim}(E) shows the electron energy spectrum. Remarkably, the resulting temperature of $\sim 7.8$ $\mathrm{MeV}$ is in good accordance with the value of 7.5 $\mathrm{MeV}$ predicted by the model developed by \textit{Pazzaglia et al.} \cite{pazzaglia2020theoretical}. About $60\%$ of the total laser energy (equal to 8.6 $\mathrm{J}$) is transferred to the electrons. We evaluate the number of accelerated electrons as the ratio between the total energy transferred to the electrons and the temperature. The result is equal to $4\times 10^{12}$, which is again in agreement with the value obtained from the model (equal to $6.4\times 10^{12}$).\\
Then, we provide the momenta distribution from the PIC to the MC to properly simulate the BS photon production in the converter, the sample and standard activation and characteristic $\gamma$-rays emission. The BS spectrum is shown in figure \ref{laser_driven_PAA_sim}(E). As expected, it has an exponential like shape and the maximum energy extends above 90 MeV. Figure \ref{laser_driven_PAA_sim}(F) reports the characteristic $\gamma$-ray spectra recorded after 3 hours of irradiation at 10 Hz repetition rate and 12 hours, 1 week and 1 month cooling times. The characteristic peaks emerge over the background. As in the case of conventional PAA, several peaks disappears for longer cooling times. This allows to avoid spectral interference and to better identify the corresponding elements. \\
Finally, exploiting the $\gamma$-ray line intensities $N_{\gamma,s}^{i}$ and $N_{\gamma,c}^{i}$ from both the sample and calibration material respectively, we can reconstruct the sample elemental concentrations $W_{s}^{i}$ originally set in the MC simulation as $W_{s}^{i} = W_{c}^{i} N_{\gamma,s}^{i}/N_{\gamma,c}^{i}$. The index $i$ refers to the different elements and $W_{c}^{i}$ are the elemental concentrations in the calibration material. The comparison between the retrieved and original elemental concentrations is shown in figure \ref{laser_driven_PAA_sim}(F). Overall, the agreement is excellent, suggesting that the proposed laser-driven source could be exploited for PAA quantitative analysis. For all the elements, the predicted concentrations slightly overestimate the actual ones. This is ascribable to the fact that the sample is placed in front of the calibration material in the MC simulation. Therefore, the photon flux seen by the calibration material is partially attenuated by the sample. This effect can be avoided exploiting a flux monitor, as already done in conventional PAA \cite{segebade2006PAA_rew}.

\section{Conclusions}
Laser-driven radiation sources are of great interest as potential multi-functional tool for elemental analysis of materials. In this work, we have shown that super-intense lasers and near-critical targets can be exploited to generate high-energy photons suitable for Photon Activation Analysis. We have proposed a theoretical framework for the identification of the optimal target and converter parameters in a wide range of laser intensities. In addition, the model can be exploited to evaluate the performances of the source in terms of activated nuclei. It is worth mentioning that our approach can be useful also to assess the potential of a laser-based photon source for other interesting applications like radioisotopes production. Last, by means of realistic simulations, we have shown that existing laser facilities are suitable for laser-driven PAA studies. Therefore, the subject of this study represents a further step toward the development of a multi-radiation platform for different materials science studies.

\section{Methods}
\subsection{Monte Carlo method}
We performed several Monte Carlo simulations exploiting the Fluka MC code \cite{ferrari2005fluka}. The fist set of simulations involved the BS production of $\gamma$-rays. The primary monoenergetic electrons are defined with the BEAM and BEAMPOS cards. On the other hand, when the input energy spectrum is non-monoenergetic, a user routine (source.f) coupled with the SOURCE card is exploited. The tungsten converter has a parallelepipedal shape. Associated to this volume, we activate the EMFCUT card. This option allows us to reduce the computational time by setting 1 MeV and 100 keV energy thresholds for the pair and photon production, respectively. By means of the USRBDX card, we retrieve both the photon energy spectra and the double differential photon spectra in energy and solid angle. All the BS simulations are performed with $10^7$ primary events per cycle and with 5 cycles.\\
As far as the simulation of the sample and standard material activation is concerned, we provide the primary $\gamma$-ray energy spectrum to the MC by sampling from the BS photon energy distribution. Since no photonuclear reactions of interest can take place below 5 MeV electron energy, the primary photons are extracted above this threshold. The sample and calibration materials are placed one in front of the other at 10 cm from the source point separated by a 1 mm gap. They are two slabs of thicknesses equal to 3 mm and infinitely extended in the orthogonal plane. Their elemental mass concentrations are reported in table \ref{composition_table}. We activate again the EMFCUT for both the sample and standard material. Then, we activate the PHOTONUC card to switch on the photonuclear reactions. To further enhance the statistical accuracy of the results, a biasing is performed by means of the LAMBIAS card. This allowed us to reduce the mean free path of photons by a factor of $10^{-3}$. The radioactive decay is carried out exploiting the RADDECAY card. We also activate the PHYSICS card with the EVAPORAT option, which describes the decay with the evaporation model considering also heavy fragment evaporation. The irradiation conditions (i.e. current of primary particles ad duration) are set by means of the IRRPROFILE card.\\
In order to acquire the characteristic $\gamma$-ray spectra collected during a certain measurement time, we exploit the DCYTIMES and USRBDX cards. Since Fluka does not allows us to automatically obtain energy spectra integrated along a certain time interval, we record the activity at many times within the overall measurement period. The instants at which we retrieve the activity are defined with the DCYTIMES card. We associate a USRBDX card to each instant in order to get the photon energy spectra emitted from the sample and standard per unit time. Then, we integrate that spectra in the time variable to obtain the signal collected during the whole measurement period. It is worth pointing out that this procedure is reliable only if the time sampling is much smaller compared to the shorter half-life for the considered activated isotopes.
\begingroup
\setlength{\tabcolsep}{5pt} 
\renewcommand{\arraystretch}{1} 
\begin{table}[h!]
\caption{Elemental mass concentrations in the sample and calibration material.}
\centering
\footnotesize
\begin{tabular}{ccccccccccccc}
\toprule
Element              & Cu    & Ca     & Ni     & Ma    & Cl    & Fe    & Pb   & Z     & Al    & Si    & Na    & Po   \\ 
Sample               & 0.404 & 0.005  & 0.001  & 0.01  & 0.012 & 0.3   & 0.08 & 0.11  & 0.02  & 0.028 & 0.02  & 0.01 \\
Calibration material & 0.33  & 0.0137 & 0.0003 & 0.006 & 0.036 & 0.025 & 0.19 & 0.074 & 0.047 & 0.018 & 0.015 & 0.02 \\
\bottomrule
\end{tabular}\label{composition_table}
\end{table}
\endgroup

\subsection{Model for the laser interaction with a near-critical plasma}
The model proposed in \cite{pazzaglia2020theoretical} allows us to describe the propagation of a super-intense laser pulse in a near-critical material and the generation of hot electrons. Here, we provide a summary of the formulas we exploited in Section \ref{sec_model_PAA_laser}. The laser pulse temporal profile is described with a $\cos^{2}$ function and a Gaussian transverse shape. To model the self-focusing effect, the evolution of the beam waist $w(r)$ along the propagation direction \textit{r} (i.e. the depth in the near-critical layer) is treated exploiting the thin-lens approximation:
\begin{equation} \label{eq:beam_waist}
\frac{w(\bar{r})}{\lambda} \approx \sqrt{\frac{1}{\pi^{2}\bar{n}} + \left(\bar{r} - \frac{w_{0}}{\lambda}\right)^2}
\end{equation}
where $\bar{r} = \sqrt{\bar{n}r/\lambda}$ is a normalized space variable, $\lambda$ is the laser wavelength and $\bar{n} = n_{nc}/\gamma_{0}n_{c}$ is a relativistic transparency factor. $\gamma_{0} = \sqrt{1+a_{0}^{2}/\wp}$ is the average Lorentz factor of the electron motion (with $\wp$ equal to 1 and 2 for circular ad linear polarization, respectively). Here, the hypotheses are that the beam waist during the propagation in the plasma $w_{m}$ is significantly smaller compared to the initial one $w_{0}$ and it keeps large compared to $\lambda$. During the propagation, the pulse heats the electrons and looses energy according to the ponderomotive scaling. The normalized energy loss can be described as:
\begin{equation} \label{eq:energy_loss}
\frac{1}{\epsilon_{p0}}\frac{d\epsilon_{p}(r)}{dr} = -2 \left(\frac{2}{\pi}\right)^{3/2}V_{2}C_{nc}\frac{1}{\tau c}\frac{n}{a_{0}n_{c}}\frac{\gamma(r)-1}{a_{0}} \left( \frac{\beta w(r)}{w_{0}}\right)^{2}
\end{equation}
where $\epsilon_{p0} = \pi^{3/2}2^{-3/2-1}m_{e}c^{2}n_{c}a_{0}^{2}w_{0}^{3-1}\tau c$ is the initial energy of the Gaussian pulse in three dimensions, $\tau$ is the field temporal duration, $V^{2}$ is the volume of a 2-dimension hypersphere with unitary radius, $C_{nc}$ is a constant accounting for the details of the electron heating, $\gamma_{r} = \sqrt{1+a(r)^{2}/\wp}$ is the local value of the Lorentz factor. $\beta$ is the ratio of the plasma channel radius to the waist assumed to be constant. To solve equation \ref{eq:energy_loss}, the pulse amplitude along the propagation length $a(r)$ is obtained from:
\begin{equation} \label{eq:local_amp}
a_{0}(r) = a_{0}\sqrt{\frac{\epsilon_p(r)/\epsilon_{p0}}{(w(r)/w_{0})^{2}}}
\end{equation}
Equations \ref{eq:local_amp} and \ref{eq:energy_loss} describe the pulse propagation in the near-critical plasma and they are solved numerically with a finite difference method. In \cite{pazzaglia2020theoretical}, the free parameters $\beta$ and $C_{nc}$ are evaluated by fitting the data obtained with 2D-PIC simulations with the model.\\
To describe the evolution of the hot electron population, it is assumed that all the energy lost by the pulse is absorbed by the electrons. This hypothesis is valid for short laser pulses (tens of fs) ad $a_{0} < 50$. The fraction of laser energy given to the electrons is 
\begin{equation} \label{eq:frac_ene_el}
\eta_{nc}(r) = \epsilon_{p}(r) / \epsilon_{0} - R_{D}
\end{equation}
where $R_{D}$ is the reflectance of the plasma. The evaluation of $R_{D}$ is not trivial, since its value depends on the considered region of the pulse. Indeed, close to the laser peak, the electrons are relativistic and the plasma is near-critical allowing the pulse propagation. On the other hand, in close proximity to tails, the electrons can be not-relativistic, resulting in an overcritical reflecting plasma. Considering the aforementioned effects, an analytical expression for $R_{D}$ in three dimensions is:
\begin{equation} \label{eq:Ref_index_3D}
R_{3D} = \mathrm{erf}(\sqrt{-2\log\bar{n}})\frac{4}{\sqrt{2\pi}}\bar{n}^{2}\sqrt{-\log\bar{n}}
\end{equation}
The number of hot electrons in the near-critical layer $N_{nc}(r)$ is given by:
\begin{equation} \label{eq:n_electrons_nc}
\frac{dN_{nc}(r)}{dr} = V_{2}n_{nc}(\beta w(r))^{2}
\end{equation}
From equation \ref{eq:frac_ene_el} and \ref{eq:n_electrons_nc} the hot electrons energy results $E_{nc}(r) = \eta_{nc}(r)\epsilon_{p0}/N_{nc}(r)$. When the pulse reaches the substrate, it generates hot electrons at the interface $r_{nc}$ with energy:
\begin{equation} \label{eq:n_electrons_inter}
E_{s}(r_{nc}) = C_{s}(\gamma(r_{nc})-1)m_{e}c^{2}
\end{equation}
where $C_{s}$ collects all the effects at the interface. As far as the absorption efficiency at the interface is concerned, it can be expressed as $\eta_{s} =N_{s}(r_{nc})E_{s}(r_{nc})/\epsilon_{p}(r_{nc}) = 0.00388 a_{0} + 0.0425$, where the coefficients of the linear relation are obtained from PIC simulations. Then, the overall electron energy $T_{e}(r_{nc})$ is obtained by combining the contributions of near-critical and substrate populations:
\begin{equation} \label{eq:overall_ele_en}
T_{e}(r_{nc}) = \frac{\eta_{s}\epsilon_{p}(r_{nc})+\eta_{nc}(r_{nc})\epsilon_{p0}}{N_{e}(r_{nc})}
\end{equation}
where:
\begin{equation} \label{eq:overall_ele_numb}
N_{e}(r_{nc}) = N_{s}(r_{nc})+N_{nc}(r_{nc})
\end{equation}
is the total number of electrons. In our work we use this model to obtain $T_{e}$ and $N_{e}$ defined in formulas \ref{eq:overall_ele_en} and \ref{eq:overall_ele_numb} for different values of $a_{0}$, $r_{nc}$ and $n_{nc}$.

\subsection{Particle-In-Cell (PIC) simulation}
The 3D Particle-In-Cell (PIC) simulation was performed with the open-source code WarpX \cite{VAY2018476}. We exploited a computational box of $\mathrm{75\lambda \times 75\lambda \times 75\lambda}$ with a spacial resolution of 20 points per $\lambda$ along all directions, where $\lambda = 800$ $\mathrm{nm}$. Time resolution was set at 98\% of the Courant Limit. The laser was linearly polarized in the simulation plane, its transverse and longitudinal profiles were Gaussian. The laser $a_{0}$ was 20.5, the waist was 4.7 $\mathrm{\mu}$m and the time duration was 30 fs. The incidence angle was $0\degree$. The target consisted of a $\mathrm{25\lambda}$ thick foil with a density of 2.92 $\mathrm{n_{c}}$. It was sampled with 4 macro-electrons and 2 macro-ions with $Z = 6$ and $A = 12$. The electron population was initialized with Maxwell-Boltzmann momentum distribution and temperature equal to 10 eV. The ion population was initialized cold. The front target-vacuum interface was at $\mathrm{25\lambda}$. The duration of the simulation was 300 fs and the total number of simulated time steps was 4640. We retrieved the electric and magnetic laser field components, electron density and macro-electrons momenta every 105 time steps. 

\printbibliography

\section*{Data availability}
The data supporting the findings of this work are available from the corresponding authors on request.

\section*{Author contribution statement}
F.M. papered the manuscript, developed the theoretical model for the comparison between conventional and laser-driven PAA, carried out the Monte Carlo simulations and contributed to the PIC simulation. D.C. contributed to the preparation of the manuscript, the model development and Monte Carlo simulation. A.F. performed the PIC simulation and revised the manuscript. M.P. conceived the project, supervised all the activities and revised the manuscript.
 
\section*{Acknowledgements}
This project has received funding from the European Research Council (ERC) under the European Union’s Horizon 2020 research and innovation programme (ENSURE grant agreement No 647554). We also acknowledge LISA and Iscra access schemes to MARCONI HPC machine at CINECA(Italy) via the project THANOS.

\section*{Competing Interests}
The authors declare no competing interests.

\end{document}